\documentclass[journal]{IEEEtran}
\ifCLASSINFOpdf
\else
\fi

\usepackage{algorithm}

\usepackage{amsmath}
\usepackage{amssymb}
\usepackage{bm}
\usepackage[pdftex]{graphicx}
\usepackage{multirow}
\usepackage{makecell}
\usepackage{upgreek}
\usepackage{newtxmath}
\usepackage{booktabs}
\usepackage{color}
\usepackage{cite}
\usepackage{soul}
\usepackage{threeparttable}

\usepackage{algorithmic}

\ifCLASSOPTIONcompsoc
  \usepackage[caption=false,font=normalsize,labelfont=sf,textfont=sf]{subfig}
\else
  \usepackage[caption=false,font=footnotesize]{subfig}
\fi

\usepackage[hidelinks]{hyperref}

\usepackage{xcolor}
\usepackage{colortbl}
\usepackage{color} 
\usepackage[normalem]{ulem} 
\usepackage[makeroom]{cancel} 
\usepackage{soul}

\definecolor{darkblue}{RGB}{5,39,175}
\newcommand{\blueText}[1]{\textcolor{black}{#1}}
\newcommand{\blueHighLight}{\color{black}}

\newcommand{\blueEnv}{\blueHighLight{}}

\begin{document}

\title{Multi-stage Moving Target Defense: A Security-enhanced D-FACTS Implementation Approach}

\author{Jiazhou~Wang,
        Jue~Tian,~ Yang~Liu,~\IEEEmembership{Member,~IEEE,}
        Xiaohong~Guan,~\IEEEmembership{Fellow,~IEEE,}
       Dong~Yang and~Ting~Liu,~\IEEEmembership{Member,~IEEE}
\thanks{Jiazhou Wang, Yang Liu, Xiaohong Guan and Ting Liu are with the Ministry
	of Education Key Lab for Intelligent Networks and Network Security, School of Cyber Science and Engineering,
	Xi’an Jiaotong University, Xi’an 710049, China (e-mail: wangjiazhou@stu.xjtu.edu.cn; yangliu@xjtu.edu.cn; xhguan@mail.xjtu.edu.cn; tingliu@mail.xjtu.edu.cn).}
\thanks{Jue Tian is with School of Computer, Xi'an University of Posts and Telecommunications, Xi’an 710061, China (e-mail: juetian@xupt.edu.cn). Corresponding author of this paper.}
\thanks{Dong Yang is with Power Infrastructure Cyber Security Research Center, Xi’an Thermal Power Research Institute Co., LTD, Xi’an 710054, China (e-mail:yangdong@tpri.com.cn).}
}

\maketitle

\begin{abstract}
In recent studies, moving target defense (MTD) has been applied to detect false data injection (FDI) attacks using distributed flexible AC transmission system (D-FACTS) devices.
\textcolor{black}{However, the inherent conflict between the security goals of MTD (i.e., detecting FDI attacks) and the economic goals of D-FACTS devices (i.e., reducing power losses) would impede the application of MTD in real systems. Moreover, the detection capabilities of existing MTDs are often insufficient. This paper proposes a multi-stage MTD (MMTD) approach to resolve these two issues by adding a group of designed security-oriented schemes before D-FACTS' economic-oriented scheme to detect FDI attacks. 
We keep these security-oriented schemes for a very short time interval and then revert to the economic-oriented scheme for the remaining time to ensure the economic requirements. We prove that a designed MMTD can significantly improve the detection capability compared to existing one-stage MTDs. We find the supremum of MMTD's detection capability and study its relationship with system topology and D-FACTS deployment.} Meanwhile, a greedy algorithm is proposed to search the MMTD strategy to reach this supremum. Simulation results show that the proposed MMTD can achieve the supremum against FDI attacks while outperforming current MTD strategies on economic indicators.


\end{abstract}


%
\IEEEpeerreviewmaketitle

\section{Introduction}\label{sec:introduction}

\IEEEPARstart{R}{ecent} studies have shown that attackers can attack the power grid by modifying measurement and control information, posing a serious threat to the power grid.   \textcolor{black}{Specifically, false data injection (FDI) attacks  can bypass the power system state estimation (SE) and bad data detection (BDD) mechanism by elaborately constructing false electrical measurement data \cite{liu2011false,6503599}}. To defend against FDI attacks, scholars have proposed many methods to protect sensors' measurements and enhance the mechanisms of data integrity checking~\cite{2010Detecting,ashok2016online}. 
{In recent years, the concept of moving target defense (MTD) has been introduced into the field of power grid security as an effective method to defend against FDI attacks~\cite{6149267}. The idea of MTD in the power grid is to bring dynamic changes to the system measurement matrix by modifying the line reactance, thus invalidating the attacker's knowledge of the measurement matrix.}

\textcolor{black}{Distributed flexible AC transmission system (D-FACTS) devices, which are widely deployed on transmission lines, have been selected by many researchers to implement the security-oriented scheme (i.e., MTD) in the power system. Specifically, we can perturb line impedance via D-FACTS devices to maximize MTD's detection capability.
However, the primary purpose of D-FACTS is to implement an economic-oriented scheme, which minimizes power losses via perturbing line impedance.
Therefore, due to the inevitable conflict between the economic-oriented scheme and the security-oriented scheme, there are two major problems when leveraging D-FACTS for MTD:
}

\blueText{1) \textbf{The operation cost with MTD strategies will increase.} In most work, D-FACTS devices are used to minimize power losses, improve the power quality and system robustness, etc. \cite{rogers2008some}.
When MTD is performed, the extra operation cost is non-ignorable \cite{liu2018reactance,2020cost}. For example, as shown in Table~\ref{tab:3buscase}, the real power losses of economic-oriented strategy in \cite{rogers2008some} are 1.167 MW. When a joint optimization strategy of economic and security goals in \cite{liu2018reactance} is applied, the real power losses will increase by 12\%.
Furthermore, if a security-oriented strategy in \cite{li2019feasibility} is applied, the real power losses will increase by 55\%. The significant increase in real power losses would prevent system operators from adopting MTD.}

\textcolor{black}{2) \textbf{The detection capabilities of existing MTD strategies are insufficient.}
The detection capability of MTD is mainly determined by three factors: system topology, D-FACTS deployment, and D-FACTS operation strategy. In general, the topology cannot be changed. In this case, the fundamental limitations of MTD's detection capabilities are twofold.
First, the sparse deployment of D-FACTS devices makes it insufficient to detect FDI attacks.
Specifically, due to the limited budget,
D-FACTS devices are usually deployed on a few critical lines. For example, only five out of 65 lines are deployed with D-FACTS devices in the 48-bus system in \cite{rogers2009power}, which can only detect 10.6\% of FDI attacks. Second, the detection capability of a single MTD is usually limited. As shown in Table~\ref{tab:3buscase}, even with full deployment, two existing  MTD strategies still suffer from undetectable attacks. Therefore, for a certain system topology and fixed D-FACTS deployment in practical applications, it is necessary to know what is the maximum detection capability and how to reach it.}

\textcolor{black}{To address the above issues, we propose a multi-stage MTD (MMTD) strategy in this paper. 
First, to resolve the conflict between security goals and economic goals, we execute different schemes with these goals sequentially within one system cycle. By keeping the execution duration of the security-oriented scheme short, we can minimize the side effects on economic performance over the entire cycle. 
Then, we prove that the superposition of a set of elaborately designed security-oriented schemes can effectively improve the detection capability. We define and demonstrate the supremum of detection capability for FDI attacks. Moreover, we propose a search algorithm based on the greedy strategy, which can reach the supremum in most cases.} Experimental results show that our strategy can achieve the maximum detection capability on five IEEE test systems (i.e., 6-bus, 14-bus, 39-bus, 57-bus, and 118-bus), while current one-stage MTDs can only achieve the maximum detection capability on the 6-bus system. The main contributions of this paper are as follows:
\begin{itemize}
\item \textcolor{black}{We propose MMTD, a security-enhanced D-FACTS implementation approach, to solve the inherent conflict between different goals when leveraging D-
FACTS for MTD.}
\item \textcolor{black}{We reveal that the system's detection capability is mainly determined by the system topology and D-FACTS deployment. We further determine the supremum of detection capability under different system conditions. We derive the minimum D-FACTS deployment condition to reach the maximum detection capability under a given topology. We also prove that existing MTD strategies, even with full D-FACTS deployment, are difficult to reach this supremum due to the topology constraint.}
\item \textcolor{black}{We propose a greedy strategy-based MMTD search algorithm to maximize the detection capability as quickly as possible. Simulation results show that the MMTD strategy outperforms current MTD strategies on both security and economic indicators.}
	
\end{itemize}

The rest of this paper is organized as follows. Preliminaries and related work are provided in Section \ref{sec:Pre}. Section \ref{sec:casestudy} proposes the basic idea of MMTD. Section \ref{sec:MMTD} proposes the metric of MMTD's detection capability and the supremum of different D-FACTS deployments. Section \ref{sec:implementation} puts forward the system model with MMTD and the implementation algorithm for maximum detection capability. Section \ref{sec:simulation} verifies the proposed strategy through numerical simulation, and Section \ref{sec:conclusion} concludes the paper.

\section{Preliminaries}\label{sec:Pre}
\blueText{In this section, we introduce the basic idea of SE, FDI attacks and MTD in the power grid, and then introduce the related work.}
\subsection{FDI Attacks against State Estimation}
In this paper, we use boldface uppercase (e.g., $ \mathbf{H} $) and lowercase (e.g., $ \mathbf{x} $) letters to indicate matrices and vectors, respectively.
\textcolor{black}{The nonlinear AC model is computationally complex and difficult to converge for large power systems \cite{liu2021converter}. Thus, a linearized DC model is widely used for state estimation, which ignores transmission line resistance and assumes identical bus voltage magnitude.}

Let $ \mathbf{z}={{({{z}_{1}},{{z}_{2}},{{z}_{3}},\ldots ,{{z}_{m}})}^{\mathsf T}}$denote the  measurement data and $ {\boldsymbol{ \thetaup }}={{({{\theta }_{1}},{{\theta }_{2}},{{\theta }_{3}},\ldots ,{{\theta }_{n}})}^{\mathsf T}} $ be the state variable ($ n+1 $ is the number of buses). In general, $ m\ge n $. In DC model, the relation between the  measurement data and the state variable is $ \mathbf{z}=\mathbf{H\boldsymbol\thetaup }+\mathbf{w} $, where $ \mathbf{H}\in {{\mathbb{R}}^{m\times n}} $ is the measurement matrix and $ \mathbf{w}\in {{\mathbb{R}}^{m}} $ is the vector of measurement noises. In general, the matrix is full rank, i.e., $rank(\mathbf{H})=n$. 
If the measurement noises are independent and identically distributed Gaussians, the system state is estimated by
$\mathbf{\hat{\boldsymbol\thetaup}}={{({{\mathbf{H}}^{\mathsf T}}{{\mathbf{R}}^{-1}}\mathbf{H})}^{-1}}{{\mathbf{H}}^{\mathsf T}}{{\mathbf{R}}^{-1}}\mathbf{z} $ \cite{abur2004power}.  $ \mathbf{R}=diag(\sigma _{1}^{-2},\sigma _{2}^{-2},\ldots ,\sigma _{m}^{-2}) $ is the covariance matrix of $ \mathbf{w} $, where ${{\sigma }_{i}}$ represents the standard deviation of measurement noise of $ {{z}_{i}} $.

\textcolor{black}{The estimation residual $ r=\Vert{{ \mathbf{z}-\mathbf{H\hat{\boldsymbol\thetaup }} }\Vert_{2}} $ refers to the deviation between the actual and estimated value of measurement vector.} 
BDD compares the residual with the pre-determined threshold $ \eta $ to determine whether the system has error data. Specifically, BDD yields a positive detection result if $r>\eta $.
Assume the attacker tampers with the system measurement as $ {{\mathbf{z}}_{a}} $ in an FDI attack, i.e., $ {{\mathbf{z}}_{a}}=\mathbf{z}+\mathbf{a}$, where $ {{\mathbf{z}}_{a}}\in {{\mathbb{R}}^{m}}$ and $\mathbf{a}\in {{\mathbb{R}}^{m}} $. The attack can bypass BDD if the attack vector $ \mathbf{a} $ satisfies $ \mathbf{a}=\mathbf{Hc} $, where $  \mathbf{c}\in {{\mathbb{R}}^{n}} $ is an arbitrary vector \cite{liu2011false}. In other words, $ \mathbf{a}\in {\text{col}(}\mathbf{H}) $, where col (·) represents the column space
of a matrix.

\subsection{Moving Target Defense}
For a transmission line $l$, we have ${{z}_{l}}=-{{b}_{l}}({{\theta }_{l,f}}-{{\theta }_{l,t}})+{{w}_{l}}$, where $ {{\theta }_{l,f}} $ and $ {{\theta }_{l,t}} $  are the state variables of the \textit{from} and \textit{to} buses of line $ l $,  respectively; $ b_l $ is the susceptance of line $ l $; $ z_l $ is the measurement value of power flow of line $ l $, and ${{w}_{l}}$ is the measurement noise of ${{z}_{l}}$.
Therefore, the row vector corresponding to line $ l $ in the measurement matrix ${\mathbf{H}}$ (denoted as ${{\mathbf{H}}_{l\cdot }}$) is \cite{enhanced}
\begin{equation}
	{{\mathbf{H}}_{l\cdot }}=[0\ ...\ 0\quad \underbrace{-{{b}_{l}}}_{l,f\text{ column}}\quad 0\ ...\ 0\quad \underbrace{{{b}_{l}}}_{l,t\text{ column}}\quad 0\ ...\ 0]
\end{equation}

\textcolor{black}{Since the measurement matrix ${\mathbf{H}}$ is a function of branch reactance, we can modify ${\mathbf{H}}$ to $\mathbf{{H}'}$ by perturbing the transmission line parameters \cite{4039419}. The defender can detect the highly structured FDI attacks $\mathbf{a}=\mathbf{Hc}$ with the new measurement matrix $\mathbf{{H}'}$.}

\begin{table*}[t]
	\centering \vspace{-22pt}
	\renewcommand{\arraystretch}{1.3}
	\caption{Detection Result of Several MTD Strategies on the 3-bus System}
	\label{tab:3buscase}%
	\begin{threeparttable}
	\begin{tabular}{c c c c c c c c }
		\toprule
		Case                    &            $ \mathbb{K}_D $                  & \multicolumn{1}{c}{$ x_1$ (p.u.) }    & \multicolumn{1}{c}{$ x_2$ (p.u.) }     & $ x_3$ (p.u.)      & \textit{DoA}                & Feasible attack         & \textcolor{black}{Real power losses (MW)}          \\ \midrule
		\multirow{2}{*}{Original Case}               &           \multirow{2}{*}{$\varnothing$ }                  & \multirow{2}{*}{0.0504} & \multirow{2}{*}{0.0572} & \multirow{2}{*}{0.0636} & \multirow{2}{*}{2}                  &        $ {[}1.1349\  1\  0{]}^{\mathsf T} $                 & \multirow{2}{*}{1.641}             \\  \cline{7-7} \specialrule{0em}{1pt}{0pt}
		               &                             &  &  & &                   &        $ {[}\text{-}1.2619\  0\  1{]}^{\mathsf T} $                 &     \\ \hline
		  \specialrule{0em}{0pt}{0.5pt}Economic Strategy \cite{rogers2008some}                    & \{1, 2, 3\}                  & \multicolumn{1}{c}{0.0606} & \multicolumn{1}{c}{0.0686} & 0.0763 & 1                  &       $ {[}\text{-}0.0769\  1.0441\  1{]}^{\mathsf T} $                  & 1.167             \\ \hline \specialrule{0em}{0pt}{0.5pt}
			 i  (MTD Strategy 1 \cite{liu2018reactance})  & \{1\}                  & \multicolumn{1}{c}{0.0605} & \multicolumn{1}{c}{0.0572} & 0.0636 & 1                  & $ {[}\text{-}1.2619\ 0\  1{]}^{\mathsf T} $ & 1.310            \\ \hline \specialrule{0em}{0pt}{0.5pt}
	ii  (MTD Strategy 2 \cite{li2019feasibility}) & \{1, 2\}                  & \multicolumn{1}{c}{0.0479} & \multicolumn{1}{c}{0.0572} & 0.0604 & 1                  & $ {[}0\  1.1119\  1{]}^{\mathsf T} $  & 1.806             \\ \hline
		\multirow{3}{*}{ iii (MMTD Strategy)} & \multirow{3}{*}{\{1, 2, 3\}} & 0.0605                      & 0.0572                      & 0.0636 & \multirow{3}{*}{0} & \multirow{3}{*}{\{\textbf{0}\}}  & \multirow{3}{*}{$ 1.167(1-\omega)+\dfrac{1.310+1.806}{2}\omega $} \\ \cline{3-5}
		&                              & 0.0479                      & 0.0572                      & 0.0604 &                    &                         &         \\ \cline{3-5}
		&                              & 0.0606                      & 0.0686                      & 0.0763 &                    &                         &         
		\\ \bottomrule 
		\vspace{-10pt}
	\end{tabular}
	\begin{tablenotes}    
        \footnotesize               
        \item[*] \textcolor{black}{$ \mathbb{K}_D $: Line set with D-FACTS services. \textit{DoA}: Dimension of the attack space.   $\omega$: The time proportions of MTD  strategy 1 and MTD  strategy 2.}        
      \end{tablenotes}            
	\end{threeparttable}
\end{table*}
Specifically, if the reactance of line $ l $, $x_l$, is modified by D-FACTS to $x'_l$, where $x'_l$ is a value within its physical limit (usually $\pm 20\%$ \cite{4039419}), the measurement matrix ${\mathbf{H}}$ becomes $\mathbf{{H}'}$. If the attack vector $\mathbf{a}=\mathbf{Hc}$ cannot be detected after MTD, there must have $\mathbf{a}\in \text{col}({{\mathbf{H}}})\bigcap \text{col}(\mathbf{{H}'})$ \cite{enhanced}. In other words, if there is no $\mathbf{{c}'}$ that satisfies $\mathbf{a}=\mathbf{Hc}=\mathbf{{H}'{c}'}$, ${\mathbf{a}}$ can be detected by BDD. In this term, the rank of the composite matrix $\mathbf{M=}\left[ \begin{matrix}
\mathbf{H} & {\mathbf{{H}'}}  \\
\end{matrix} \right]$ can be used to characterize the detection capability of MTD, where $\mathbf{M}\in {{\mathbb{R}}^{m\times 2n}}$. The detection capability increases as the rank of $ \textbf{M} $ increases. In particular, an MTD can detect any FDI attack when $rank\left( \mathbf{M} \right)\mathbf{=}2n$, which is called a complete MTD \cite{enhanced}. However, a complete MTD is difficult to achieve because the system must satisfy $m\ge 2n$. In this paper, we give a new condition of complete MTD in Section \ref{sec:MMTD} to make it more applicable.

\subsection{Related Work}

As typical cases of FDI attacks, several recent security incidents, such as Stuxnet, BlackEnergy3, have brought serious threats to power grids \cite{liu2020optimal}. FDI attacks assume that the attacker can obtain the power system's topology and real-time configuration information and tamper with the electricity meter's measurement value \cite{liu2011false}. 
Since the power system has many remote unattended infrastructures, it is impossible to protect all transmission lines \cite{li2019feasibility}. In general, FDI attacks pose a significant threat to the power system.

\begin{figure}[t]\vspace{-10pt}
	\centering
	\includegraphics[width=0.35\linewidth]{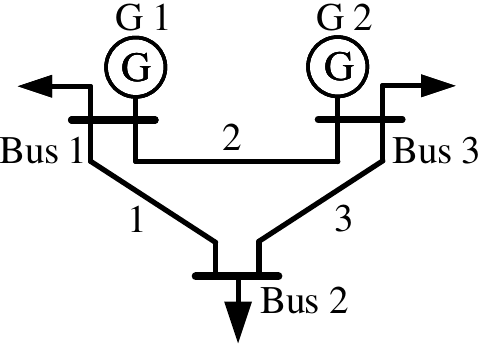}\vspace{-5pt}
	\caption{A 3-bus system.}
	\label{fig:3bus}\vspace{-15pt}
\end{figure}

MTD is considered to be an effective method to defend against FDI attacks. Rahman \textit{et al.} proposed a 
random MTD strategy in \cite{rahman2014moving}. The defender randomly selects a set of transmission lines and randomly changes the transmission line reactance, making the attacker unable to obtain the latest system information. Tian \textit{et al.}  defined the stealthiness of MTD in \cite{hidden}, and proved the mutual exclusion of the completeness and stealthiness in \cite{enhanced}. Liu \textit{et al.} discussed the D-FACTS deployment problem of MTD in practical application~\cite{liu2020optimal}.
Lakshminarayana \textit{et al.} \cite{lakshminarayana2019moving} identified the set of lines whose reactances must be perturbed at the operational time based on a game-theoretic approach and discussed the D-FACTS deployment problem from a security point of view only.
Li \textit{et al.} \cite{li2019feasibility} discussed the feasibility and limitations of MTD. Lakshminarayana \textit{et al.} \cite{2020cost} used the Lebesgue measure between the measurement matrices before and after MTD to indicate the MTD's effectiveness. Liu \textit{et al.} \cite{liu2018reactance} used the rank of a composite measurement matrix to characterize the effectiveness of MTD and proposed a heuristic algorithm that maximizes the rank of the composite matrix and minimizes system losses. \textcolor{black}{Zhang et al. proved the condition that MTD can detect all attacks in \cite{zhang2020Analysis}, and proposed two countermeasures to address the attacks that can be constructed with no branch parameters knowledge in \cite{zhang2021Zero}. They further developed a heuristic algorithm to compute a near-optimal deployment scheme of D-FACTS devices and proposed a coordination strategy for consecutive perturbation schemes in \cite{zhang2021cycle}.}
However, the detection capabilities of the existing MTD studies are insufficient, and the corresponding relationship between the implementation scheme and the protected lines is not disclosed. 

The application of D-FACTS is originally intended to improve the system's economic performance and robustness \cite{rogers2008some}. The results in \cite{2020cost} show that an improvement in the detection effectiveness of an MTD might increase the system operating cost. A heuristic algorithm was proposed in \cite{2020cost} to seek joint optimization of economic goals and security performance. Because the security performance of MTD is related to the difference between the two measurement matrices before and after MTD, the joint optimization problem is often a rank-constrained problem, which has been proven to be an NP-hard problem and difficult to solve \cite{sun2017rank}.

With the aforementioned issues, existing MTD strategies are still not practical. In this paper, we will propose an MMTD strategy to strike a balance between the economic goals for power grids and the security goals for MTD.
\begin{figure}[t]\vspace{-10pt}
	\centering
	\includegraphics[width=0.9\linewidth]{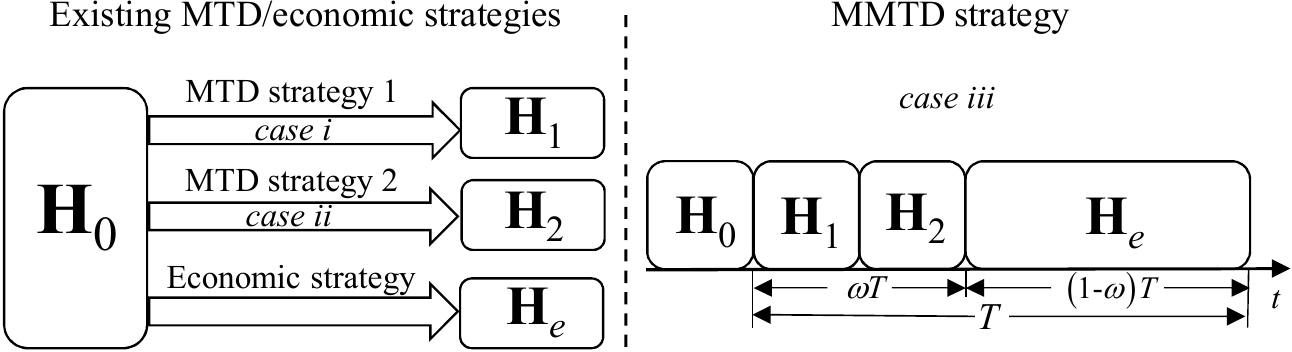}\vspace{-7pt}
	\caption{\textcolor{black}{Existing MTD/economic strategies versus the MMTD strategy.}}
	\label{fig:framework}\vspace{-10pt}
\end{figure}
\section{Basic Idea of MMTD}\label{sec:casestudy}
In this section, we use a minimal case to illustrate the issues of existing MTD strategies and show the framework of MMTD strategy.
\subsection{Case Study}
This section shows a small case of existing MTD solutions. Fig. \ref{fig:3bus} shows a 3-bus system where bus 1 is the reference bus. The original load
and generation profile is $ Pd_1 $ = 50 MW, $ Pd_2 $ = 170 MW, $ Pd_3 $ = 280 MW, $ Pg_1 $= 182 MW, $ Pg_3 $ = 318 MW, where $ Pd_j $ and  $ Pg_j $ denote the active load and the generator's power output on bus $ j $,   respectively. The line reactance profile is $ x_1 $ = 0.0504, $ x_2 $ = 0.0572, $ x_3 $ = 0.0636. \textcolor{black}{The measurement matrix is $\mathbf{{H}} = \left[ \begin{smallmatrix}
	-19.84 & 0 & 15.72\\0 & -17.48 &-15.72
\end{smallmatrix}  \right]^{\mathsf T}$. Then, the system state can be derived as $\boldsymbol{\thetaup }={{\left[ \begin{matrix}
			-3.10 & -0.81  \\
\end{matrix} \right]}^{\mathsf T}}$.} In the noiseless environment, the measurement value of the power flow $\mathbf{z}={{\left[ \begin{matrix}
			107.26 & 24.74 & -62.74  \\
\end{matrix} \right]}^{\mathsf T}}$, and the threshold is zero. As shown in Table \ref{tab:3buscase}, the economic strategy, MTD strategy 1 in case $i$ and MTD strategy 2 in case $ii$ are obtained from \cite{rogers2008some}, \cite{liu2018reactance} and \cite{li2019feasibility}, respectively. Fig. \ref{fig:framework} shows the process of the four cases, where case $iii$ is an example of MMTD strategy which is to execute MTD strategy 1 and MTD strategy 2 in turn and return to the economic-oriented scheme at last. \blueText{The time interval of the entire cycle of case $iii$ is $T$, the sum of the proportions of MTD strategy 1 and MTD strategy 2 in the cycle are $\omega$, and the proportion of the economic strategy is $1-\omega$.
According to Table \ref{tab:3buscase}, the system is still vulnerable to FDI attacks under the economic strategy, case $i$ (i.e., MTD strategy 1), or case $ii$ (i.e., MTD strategy 2), while all attacks can be detected after case $iii$ (i.e., the MMTD strategy).}

\textcolor{black}{As shown in Table \ref{tab:3buscase}, the real power losses of current MTD strategies (i.e., case $i$ and case $ii$) are much higher than that of the economic strategy, which prevents system operators from adopting MTD strategies. 
In contrast, when $\omega$ is small, the  power losses of the cycle are very close to the economic strategy, which can solve the conflict between MTD's security goals and D-FACTS' economic goals.}

\subsection{Framework of MMTD}

This section shows the framework of MMTD and the issues that need to be solved. As shown in Fig. \ref{fig:framework}, ${{\mathbf{H}}_{0}}$ is the original measurement matrix and ${{\mathbf{H}}_{e}}$ is the measurement matrix corresponding to the economic-oriented scheme. We add several security-oriented schemes ${{\mathbf{H}}_{1}},{{\mathbf{H}}_{2}},\ldots ,{{\mathbf{H}}_{k}}$ between ${{\mathbf{H}}_{0}}$ and ${{\mathbf{H}}_{e}}$, instead of modifying the economic-oriented scheme ${{\mathbf{H}}_{e}}$. 

In conclusion, the MMTD strategy can solve the conflict of different goals in taking D-FACTS to implement MTD and improve the detection capability of existing MTDs. Next, we discuss several key issues that need to be solved:

1)	How to measure the detection capability of an MTD strategy? What is the supremum of the detection capability?

2)	What are the deployment conditions of D-FACTS to achieve the supremum? What are the implementation strategies to achieve this supremum?

\section{Detection Capability of MTD}\label{sec:MMTD}
In this section, we propose a metric for MTD's detection capability and calculate the supremum of detection capability. Subsequently, we reveal the relationship between the supremum and system topology and D-FACTS deployment.

\subsection{The Metric for Detection Capability}
This section analyzes the detection capability of MMTD from the perspective of attack space and gives the corresponding metric.
\newtheorem{corollary}{\textbf {Corollary}}
\newtheorem{definition}{\textbf {Definition}}
\newtheorem{lemma}{\textbf {Lemma}}
\begin{definition} \label{Defin1}
	When there are $ k $ MTD schemes ${{\mathbf{H}}_{1}},{{\mathbf{H}}_{2}},\ldots ,{{\mathbf{H}}_{k}}$ \blueText{within} an MMTD operation period, if an attack $\mathbf{a}$ cannot be detected in the whole MMTD period, then the attack space is $\text{col}({{\mathbf{H}}_{0}})\bigcap \text{col}({{\mathbf{H}}_{1}})\bigcap \text{col}({{\mathbf{H}}_{2}})\bigcap \ldots \bigcap \text{col}({{\mathbf{H}}_{k}})$. The dimension of the attack space is called \textit{DoA}.  
\end{definition}

\begin{figure}[t]\vspace{-4pt}
	\centering
	\includegraphics[width=0.8\linewidth]{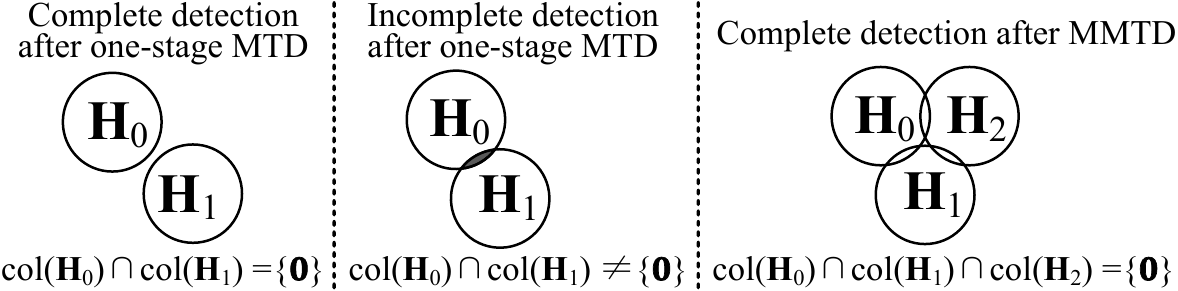} \vspace{-5pt}
	\caption{\textcolor{black}{Attack space after different MTD strategies in different systems.}}
	\label{fig:Aspace} \vspace{-16pt}
\end{figure}

For an FDI attack $\mathbf{a=}{{\mathbf{H}}_{0}}\mathbf{c}$, the attack space of $\mathbf{a}$ is the column space of the measurement matrix ${{\mathbf{H}}_{0}}$, i.e., $\mathbf{a}\in \text{col(}{{\mathbf{H}}_{\text{0}}})$. As shown in Fig. \ref{fig:Aspace}, within an MMTD operation period, the attack space gradually decreases with the increase in the number of MTD executions. In particular, if the column space intersection of multiple measurement matrices is $ \{\mathbf{0}\} $, then all attacks can be detected after the MMTD  period. From Table \ref{tab:3buscase}, the intersection of the original case and case \textit{i} and case \textit{ii} is $ \{\mathbf{0}\} $. Then all attacks can be detected after case \textit{iii}.

We then analyze FDI attacks from the perspective of graph theory. A grid topology with $ n+1 $ buses (with one reference bus) and $ m $ branches can be denoted by a graph $  \mathcal G= (\mathcal{V}, \mathcal{E}) $ with $ n+1 $ vertices and $ m $ edges, where $ \mathcal{V} $ and $ \mathcal{E} $ are the vertex set and the edge set of $ \mathcal G  $, respectively \cite{diestel2017basics}. The kernel of the incidence matrix  is the cycle space, which is spanned by all the cycles in $ \mathcal G $. \textcolor{black}{The image of the incidence matrix is the cut space, which is spanned by all the cuts in $ \mathcal G $ \cite{diestel2017basics}.} The dimensions of the cycle space and the cut space are $ m-n $ and $ n $, respectively. A set of maximally independent bases of the cycle space and the cut space are called a set of circuit basis and cut basis, respectively \cite{diestel2017basics}.

According to the definition of $ \mathbf{H} $, $ \mathbf{H} $ is a form of system incidence matrix. Suppose ${{\mathbf{s}}_{1}}\text{ }{{\mathbf{s}}_{2}}\ldots \text{ }{{\mathbf{s}}_{n}}$ are the $ n $ independent basic cuts of the system, and $\mathbf{S}={{\left[ {{\mathbf{s}}_{1}}\text{ }{{\mathbf{s}}_{2}}\ldots \text{ }{{\mathbf{s}}_{n}} \right]}^{\mathsf T}}$ is the cut basis matrix generated by $ \mathbf{H} $. 
Because the image of incidence matrix is the cut space \cite{diestel2017basics}, $ \mathbf{H} $ and $ \mathbf{S} $ have the same column space. Then the necessary condition for FDI attacks is that the attacked lines must be the union of the system cuts.

From $\mathbf{z=H\boldsymbol{ \thetaup } }$, we have ${{\mathbf{H}}^{\mathbf{+}}}\mathbf{z=\boldsymbol{ \thetaup } }$, where ${{\mathbf{H}}^{+}}={{({{\mathbf{H}}^{\mathsf T}}{{\mathbf{R}}^{-1}}\mathbf{H})}^{-1}}{{\mathbf{H}}^{\mathsf T}}{{\mathbf{R}}^{-1}}$ is the pseudo-inverse of $ \mathbf{H} $. Then the null space of ${{\mathbf{H}}^{\mathbf{+}}}$ is the cycle space of the system. According to the definition of ${{\mathbf{H}}^{\mathbf{+}}}$, ${{\mathbf{H}}^{\mathsf T}}$ and ${{\mathbf{H}}^{\mathbf{+}}}$ have the same null space. For ${{\mathbf{H}}^{\mathsf T}}\in {{\mathbb{R}}^{n\times m}}$, there are $ m-n $ basis in the null space of ${{\mathbf{H}}^{\mathsf T}}$. Let ${{\mathbf{f}}_{1}},{{\mathbf{f}}_{2}},\ldots ,{{\mathbf{f}}_{m-n}}$ be a set of basis of the null space of ${{\mathbf{H}}^{\mathsf T}}$, and $\mathbf{F}={{\left[ {{\mathbf{f}}_{1}}\text{ }{{\mathbf{f}}_{2}}\ldots \text{ }{{\mathbf{f}}_{m-n}} \right]}^{\mathsf T}}$ is the corresponding circuit basis matrix, where $\mathbf{F}\in {{\mathbb{R}}^{(m-n)\times m}}$. Then we have ${{\mathbf{H}}^{\mathsf T}}{{\mathbf{F}}^{\mathsf T}}\mathbf{=0}$, i.e., $\mathbf{FH=0}$. According to the definition of the circuit basis, the row vectors of $ \mathbf{F} $ are a set of circuit basis of the system, where each row vector of $ \mathbf{F} $ corresponds to a fundamental circle. We can further get $\mathbf{Fz=FH\boldsymbol{ \thetaup } }=\mathbf{0}$. In other words, any attack that satisfies $\mathbf{a=Hc}$ also satisfies $\mathbf{Fa}\text{=}\mathbf{0}$. 
The column space of $ \mathbf{H} $ is the null space of $ \mathbf{F} $.

\textcolor{black}{Let ${{\mathbf{F}}_{0}}$ be the circuit basis matrix corresponding to the initial measurement matrix ${{\mathbf{H}}_{0}}$, and ${{\mathbf{F}}_{k}}$ be the circuit basis matrix ${{\mathbf{F}}_{k}}$ for the $k$-th MTD, corresponding to ${{\mathbf{H}}_{k}}$. Let $\mathbf{L}$ be the composite matrix which is composed of multiple $\mathbf{F}$, and   ${{\mathbf{L}}_{k}}={{\left[ \begin{matrix}
			\mathbf{F}_{0}^{\mathsf T} & \mathbf{F}_{1}^{\mathsf T} & \mathbf{F}_{2}^{\mathsf T} & \begin{matrix}
				\ldots  & \mathbf{F}_{k}^{\mathsf T}  \\
			\end{matrix}  \\
		\end{matrix} \right]}^{\mathsf T}}$ be the composite matrix obtained by  $ k $ times MTD, where ${{\mathbf{L}}_{k}}\in {{\mathbb{R}}^{((k+1)(m-n))\times m}}$. Then, we can get Lemma \ref{lemma:nullspace}	 as follows}:
\begin{lemma}\label{lemma:nullspace}	
$\text{col}({{\mathbf{H}}_{0}})\bigcap \ldots \bigcap \text{col}({{\mathbf{H}}_{k}})\text{ = ker(}{{\mathbf{L}}_{k}}\text{)}$.
\end{lemma}
\begin{IEEEproof}
	\textcolor{black}{If an attack  $ \textbf{a} \in \text{col}({{\mathbf{H}}_{0}})\bigcap \ldots \bigcap \text{col}({{\mathbf{H}}_{k}})$, $ \textbf{a} $ satisfies $ \mathbf{F}_{0}\textbf{a}=\mathbf{F}_{1}\textbf{a} = \ldots =\mathbf{F}_{k}\textbf{a} = \textbf{0} $, $\mathbf{L}_{k}\textbf{a} = \textbf{0}$, i.e., $ \textbf{a} \in \text {ker}(\textbf{L}_k) $. Similarly, if $ \textbf{a} \in \text {ker}(\textbf{L}_k) $, then  $\mathbf{L}_{k}\textbf{a} = \textbf{0}$. We further get $ \mathbf{F}_{0}\textbf{a}=\mathbf{F}_{1}\textbf{a} = \ldots =\mathbf{F}_{k}\textbf{a} = \textbf{0}$, i.e., $ \textbf{a} \in \text{col}({{\mathbf{H}}_{0}})\bigcap \ldots \bigcap \text{col}({{\mathbf{H}}_{k}})$. Therefore, $\text{col}({{\mathbf{H}}_{0}})\bigcap \ldots \bigcap \text{col}({{\mathbf{H}}_{k}})\text{ = ker(}{{\mathbf{L}}_{k}}\text{)}$.}
\end{IEEEproof}

\textcolor{black}{Lemma \ref{lemma:nullspace} transforms the intersection of multiple matrices' column spaces into the null space of one matrix. By observing the dimension of the null space of \textbf{L}, we can accurately evaluate the detection capability of an MMTD scheme.}

The practical significance of $\mathbf{Fz=0}$ is analyzed as follows. The $ i $-th row of $ \mathbf{F} $ represents a fundamental cycle ${{C}_{i}}$ of the system, where $i=1,2,\ldots ,m-n$. The $ l $-th column of $ \mathbf{F} $ represents line $ l $ of the system, where $l=1,2,\ldots ,m$. 
We have $\left| {{\mathbf{F}}_{il}} \right|=\left\{ \begin{aligned}
	 &0,&l\notin {{C}_{i}} \\ 
	 &{{x}_{l}},&l\in {{C}_{i}} \\ 
\end{aligned} \right.$. 
We further divide $\mathbf{F}$ into $\mathbf{F}=\mathbf{GX}$, where $\mathbf{G}\in {{\mathbb{R}}^{(m-n)\times m}}$ is the loop matrix of the system, which represents the topological relationship between the line and the basic cycle, and $ {{\mathbf{G}}_{il}}=\left\{ \begin{aligned}
    & 0,&l\notin {{C}_{i}} \\ 
	& \pm 1,&l\in {{C}_{i}} \\ 
\end{aligned} \right.$.
Specially, ${{\mathbf{G}}_{il}}=\text{1}$ when $l\in {{C}_{i}}$ and the line flow in $ l $ is in the same direction as ${{C}_{i}}$;  ${{\mathbf{G}}_{il}}=-1$ when $l\in {{C}_{i}}$ and the line flow in $ l $ is in the inverse direction as ${{C}_{i}}$. $ \mathbf{X}=diag({x}_{1},{x}_{2},\ldots ,{x}_{m}) $ is a diagonal matrix composed of all line reactances of the system. In a noiseless environment, for line $ l $, we have  \begin{equation}\label{lineflow}
	{{z}_{l}}=-{{b}_{l}}({{\theta }_{l,f}}-{{\theta }_{l,t}})=\frac{{{\theta }_{l,f}}-{{\theta }_{l,t}}}{{{x}_{l}}}
\end{equation}
i.e.,  ${{x}_{l}}{{z}_{l}}={{\theta }_{l,f}}-{{\theta }_{l,t}}$. Then, $\sum\limits_{l=1}^{m}{{{x}_{l}}{{z}_{l}}}=\sum\limits_{l=1}^{m}{\left( {{\theta }_{l,f}}-{{\theta }_{l,t}} \right)}, \  l\in {{C}_{i}}$. When the power flow of all lines takes the same direction, we have $\sum\limits_{l=1}^{m}{{{\mathbf{F}}_{il}}{{z}_{l}}}=0$. In other words, the algebraic sum of the potential difference across all components along a closed loop is equal to zero, which obeys the Kirchhoff's voltage law.

According to Lemma \ref{lemma:nullspace}, Lemma \ref{lemma:rank} can be obtained as follows:
\begin{lemma} \label{lemma:rank}
	$rank({{\mathbf{L}}_{k}})+\dim\left( \text{col}({{\mathbf{H}}_{0}})\bigcap \ldots \bigcap \text{col}({{\mathbf{H}}_{k}}) \right)=m$
\end{lemma}
\begin{IEEEproof}
	As $\text{col}({{\mathbf{H}}_{0}})\bigcap \ldots \bigcap \text{col}({{\mathbf{H}}_{k}})\text{ = ker(}{{\mathbf{L}}_{k}}\text{)}$, and there are $ m $ columns in ${{\mathbf{L}}_{k}}$, we can further get   $rank({{\mathbf{L}}_{k}})+\dim\left( \text{col}({{\mathbf{H}}_{0}})\bigcap \ldots \bigcap \text{col}({{\mathbf{H}}_{k}}) \right)=m$.
\end{IEEEproof}

Obviously, the initial \textit{DoA} for any system before MTD is $ n $. The minimum \textit{DoA} after a one-stage MTD is $\max \left\{ 2n-m,0 \right\}$. Specially, the result is zero when the system satisfies $m\ge 2n$ \cite{enhanced}. We can use \textit{DoA}/$ n $ to evaluate the detection capability of an MTD strategy, where a smaller \textit{DoA}/$ n $ indicates a better detection capability.

\subsection{Detection Capability's Supremum under Full D-FACTS Deployment}
The maximum detection capability of MTD is mainly determined by the system topology and the deployment of D-FACTS devices. In this section, we consider the full deployment scenario where all lines are deployed with D-FACTS devices. With this assumption, we can study the impact of specific topology on the supremum of detection capability. We also redefine the completeness of MMTD in this section.
\begin{definition}
If the attack space of a system can be reduced to $\{\mathbf{0}\}$ by MMTD, the system is a complete system. Otherwise, the system is an incomplete system and its minimum \textit{DoA} is called degree of incompleteness (\textit{DoI}).
\end{definition}

\textcolor{black}{According to the existing definition of complete MTD, $m\ge 2n$ is the necessary condition of a complete MTD \cite{enhanced}. However, when $\dim\left( \text{col}({{\mathbf{H}}_{0}})\bigcap \ldots \bigcap \text{col}({{\mathbf{H}}_{k}}) \right)=\text{0}$, any FDI attack can be detected with MMTD.  From Table \ref{tab:3buscase}, the 3-bus system does not meet $m\ge 2n$ but can detect any attack with MMTD.}

To illustrate the topological characteristics of complete and incomplete systems, the concept of the single-line cut is defined in Definition \ref{def:scut}.
\begin{definition}\label{def:scut}
A cut with only one line is a single-line cut.
\end{definition}

\begin{figure}[t]\vspace{-5pt}
	\centering
	\includegraphics[width=0.5\linewidth]{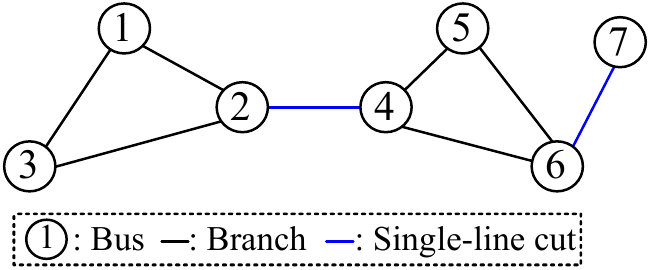} \vspace{-5pt}
	\caption{\textcolor{black}{Single-line cuts in the system.}}
	\label{fig:Scut} \vspace{-15pt}
\end{figure}

\textcolor{black}{Fig. \ref{fig:Scut} shows two forms of the single-line cut. Specially, any attack on a single-line cut ${{l}_{sc}}$ can never be detected and the MTD on ${{l}_{sc}}$ is ineffective. Assume a system can be divided into two subsystems Q1 and Q2 by disconnecting along the cut ${{l}_{sc}}$. When the system load is stable, the power flow on ${{l}_{sc}}$ is always equal to the unbalance between the subsystems Q1 and Q2. In this case, the relation always holds no matter how the parameters of ${{l}_{sc}}$ are modified.} The relationship between the single-line cut and the complete system is illustrated in Theorem \ref{Theorem:complete}.

\newtheorem{theorem}{\textbf {Theorem}}
\begin{theorem} \label{Theorem:complete}
	 A system is a complete system if and only if there is no single-line cut in the system.
\end{theorem}

\begin{IEEEproof}
(Sufficiency): According to Lemma \ref{lemma:rank}, if the system has no single-line cut, the sufficient condition for the system to be a complete system is $\exists {{\mathbf{L}}_{k}}$ where $rank({{\mathbf{L}}_{k}})=m$. In the $ k $-th MTD, we have ${{\mathbf{F}}_{k}}=\mathbf{G}{{\mathbf{X}}_{k}}$, where $ \mathbf{X}_k=diag({x}_{k,1}, {x}_{k,2}, \ldots ,{x}_{k,m}) $ is the diagonal matrix composed of all line reactances in the system at the $ k $-th MTD.  ${{\mathbf{F}}_{k}}=\left[ \begin{matrix}
	{{\mathbf{F}}_{k,\cdot 1}} & {{\mathbf{F}}_{k,\cdot 2}} & \ldots  & {{\mathbf{F}}_{k,\cdot m}}  \\
\end{matrix} \right]$,  $ \mathbf{G}=\left[ \begin{matrix}
	{{\mathbf{G}}_{\cdot 1}} & {{\mathbf{G}}_{\cdot 2}} & \ldots  & {{\mathbf{G}}_{\cdot m}}  \\
\end{matrix} \right]$, where ${{\mathbf{F}}_{k,\cdot l}}$ and  ${{\mathbf{G}}_{\cdot l}}$, are the $ l $-th column vectors of ${{\mathbf{F}}_{k}}$, and $\mathbf{G}$,  respectively. Then ${{\mathbf{F}}_{k,\cdot l}}={{x}_{k,l}}{{\mathbf{G}}_{\cdot l}}$. Consider the simplest case of MTD, where only the parameter of line $ k $ is modified at $ k $-th MTD, i.e.,
\begin{equation}\notag
	{{x}_{k,l}}=\left\{ \begin{matrix}
		(1+{{\delta }_{k}}){{x}_{0,l}}, &l = k \\ 
		{{x}_{0,l}}, &l\ne k\\  
	\end{matrix}\right.
\end{equation}
where ${{\delta }_{k}}\ne 0$. Then the composite matrix \textbf{L} after $ m $ times MTD is 
$${{\mathbf{L}}_{m}}=\left[ \begin{matrix}
	{{\mathbf{F}}_{\text{0}}}  \\
	{{\mathbf{F}}_{\text{1}}}  \\
	\vdots   \\
	{{\mathbf{F}}_{m}}  \\
\end{matrix} \right]=\left[ \begin{matrix}
	\mathbf{G} & {} & {} & {}  \\
	{} & \mathbf{G} & {} & {}  \\
	{} & {} & \ddots  & {}  \\
	{} & {} & {} & \mathbf{G}  \\
\end{matrix} \right]\left[ \begin{matrix}
	{{\mathbf{X}}_{0}}  \\
	{{\mathbf{X}}_{1}}  \\
	\vdots   \\
	{{\mathbf{X}}_{m}}  \\
\end{matrix} \right].$$

${{\mathbf{X}}_{k}}$ can be written as $\left[ \begin{matrix}
	{{x}_{k,1}}{{\mathbf{e}}_{1}} & {{x}_{k,2}}{{\mathbf{e}}_{2}} & \cdots  & {{x}_{k,m}}{{\mathbf{e}}_{m}}  \\
\end{matrix} \right]$, where ${{\mathbf{e}}_{\text{1}}}\text{=}{{\left[ \begin{matrix}
			\text{1} & \text{0} & \ldots  & \text{0}  \\
		\end{matrix} \right]}^{\mathsf T}},\ldots ,{{\mathbf{e}}_{m}}\text{=}{{\left[ \begin{matrix}
			\text{0} & \text{0} & \ldots  & \text{1}  \\
		\end{matrix} \right]}^{\mathsf T}}$ are the unit basis vectors of the $ m $-dimensional space.
Then ${{\mathbf{X}}_{k}}={{\mathbf{X}}_{0}}+\left[ \begin{matrix}
	\mathbf{0} & \begin{matrix}
		\ldots   \\
	\end{matrix} & {{\delta }_{k}}{{x}_{0,k}}{{\mathbf{e}}_{k}} & \begin{matrix}
		\ldots  & \mathbf{0}  \\
	\end{matrix}  \\
\end{matrix} \right]$. Construct ${{\mathbf{X}}^{*}}$  by 
$${{\mathbf{X}}^{*}}=\left[ \begin{matrix}
	{{\mathbf{X}}_{0}}  \\
	{{\mathbf{X}}_{1}}  \\
	\vdots   \\
	{{\mathbf{X}}_{m}}  \\
\end{matrix} \right]-\left[ \begin{matrix}
	{{\mathbf{0}}_{m\times m}}  \\
	{{\mathbf{X}}_{0}}  \\
	\vdots   \\
	{{\mathbf{X}}_{0}}  \\
\end{matrix} \right]
=\left[ \begin{matrix}
	{{x}_{0,1}}{{\mathbf{e}}_{1}} & {{x}_{0,2}}{{\mathbf{e}}_{2}} & \ldots  & {{x}_{0,m}}{{\mathbf{e}}_{m}}  \\
	{{\delta }_{1}}{{x}_{0,1}}{{\mathbf{e}}_{\text{1}}} & {} & {} & {}  \\
	{} & {{\delta }_{2}}{{x}_{0,2}}{{\mathbf{e}}_{2}} & {} & {}  \\
	{} & {} & \ddots  & {}  \\
	{} & {} & {} & {{\delta }_{m}}{{x}_{0,m}}{{\mathbf{e}}_{m}}  \\
\end{matrix} \right].$$
Denote the diagonal matrix expanded by $ m $ system loop matrices by $ {\mathbf{G}}^{*}=diag(\mathbf{G},\mathbf{G},\ldots ,\mathbf{G})$, then
 $${{\mathbf{G}}^{*}}{{\mathbf{X}}^{*}}=\left[ \begin{matrix}
	{{\mathbf{F}}_{0,\cdot 1}} & {{\mathbf{F}}_{0,\cdot 2}} & \ldots  & {{\mathbf{F}}_{0,\cdot m}}  \\
	{{\delta }_{1}}{{\mathbf{F}}_{0,\cdot 1}} & {} & {} & {}  \\
	{} & {{\delta }_{2}}{{\mathbf{F}}_{0,\cdot 2}} & {} & {}  \\
	{} & {} & \ddots  & {}  \\
	{} & {} & {} & {{\delta }_{m}}{{\mathbf{F}}_{0,\cdot m}}  \\
\end{matrix} \right].$$

When the system has no single-line cut, there is ${{m}_{sc}} \text{ = 0}$, and every column of the circuit basis matrix has non-zero elements, then $rank\left( {{\mathbf{G}}^{*}}{{\mathbf{X}}^{*}} \right)=m$, i.e., $rank\left( {{\mathbf{L}}_{m}} \right)=m$. The system is a complete system.

Necessity: Because a single-line cut does not belong to any cycle, the corresponding column of single-line cuts in the circuit basis matrix only have zero-elements. Suppose a system has ${{m}_{sc}}$ single-line cuts, and the set of all single-line cuts is ${{\mathbb{K}}_{sc}}$. Then, for an attack $\mathbf{a}=\left[ \begin{matrix}
	{{a}_{1}} & {{a}_{2}} & \ldots  & {{a}_{m}}  \\
\end{matrix} \right]$, $\mathbf{La=0}$ always holds when the attack satisfies ${{a}_{l}}=0$, $\forall l\notin {{\mathbb{K}}_{sc}}$, i.e., attacks on single-line cuts are always undetectable. 
\end{IEEEproof}
The number of single-line cuts is equal to \textit{DoI} of the system. When there are ${{m}_{sc}}$ single-line cuts in the system, there are always ${{m}_{sc}}$ columns of ${{\mathbf{L}}}$ only have zero-elements, then the \textit{DoI} of the system is ${{m}_{sc}}$. We can further get the condition to achieve the topology-determined supremum of detection capability as Corollary \ref{maxrank}.
\begin{corollary}\label{maxrank}
	The condition to achieve the topology-determined supremum of detection capability is: $rank(\mathbf{L})=m-{{m}_{sc}}$.
\end{corollary}

\vspace{-10pt}\subsection{D-FACTS Deployment Condition for Topology-determined Supremum}

In this section, we will seek the minimum deployment condition to reach the maximum detection capability under given topologies.
The complete system is discussed at first. Let ${{\mathbb{K}}_{D}}$ be the set of lines with D-FACTS, Theorem \ref{Theorem:deployment} can be derived as follows.

\begin{theorem} \label{Theorem:deployment}
	A complete system can detect any FDI attack if and only if ${{\mathbb{K}}_{D}}$ contains at least one spanning tree of the system.
\end{theorem}

\begin{IEEEproof}
	(Sufficiency): According to Lemma \ref{lemma:rank}, to maximize the detection capability is to minimize the \textit{DoA}. For a complete system, every column of the circuit basis matrix has at least one non-zero element. \textcolor{black}{Each row of the loop matrix $\mathbf{G}$ is formed by adding an edge to the spanning tree $ \mathbb{T} $, so the circle corresponding to each row has a unique edge. These edges belong to the cotree $\bar{\mathbb{T}}$ corresponding to the tree $\mathbb{T}$ \cite{diestel2017basics}.} 
	
	The loop matrix $\mathbf{G}$ can be written as $\mathbf{G}=\left[ \begin{matrix}
		{{\mathbf{G}}_{{\bar{\mathbb{T}}}}} & {{\mathbf{G}}_{\mathbb{T}}}  \\
	\end{matrix} \right]$, where ${{\mathbf{G}}_{{\bar{\mathbb{T}}}}}\in {{\mathbb{R}}^{(m-n)\times (m-n)}}$ is a diagonal matrix representing the cotree line part and ${{\mathbf{G}}_{\mathbb{T}}}\in {{\mathbb{R}}^{(m-n)\times n}}$ is the part of the tree lines.
	Suppose ${{\mathbb{T}}_{k}}$ is a line in $ \mathbb{T} $,  $k=1,\ldots ,n$. Suppose only the parameters of the line ${{\mathbb{T}}_{k}}$ is modified in the $ k $-th MTD as \begin{equation}\notag 
		{{x}_{k,l}}=\left\{ \begin{matrix}
			(1+{{\delta }_{k}}){{x}_{0,l}}, &l \text{=} {{\mathbb{T}}_{k}} \\ 
			{{x}_{0,l}}, &l\ne {{\mathbb{T}}_{k}}\\ 
		\end{matrix} \right.
	\end{equation} 
where ${{\delta }_{k}}\ne 0$. Then a row vector $\left[ \begin{matrix}
		0 & \ldots  & {{\delta }_{k}}{{x}_{0,{{\mathbb{T}}_{k}}}} & \ldots  & 0  \\
	\end{matrix} \right]$ can be obtained by  ${{\mathbf{F}}_{k}}-{{\mathbf{F}}_{0}}$. In the same way, after $ n $ times of MTD, we get
	$$\left[ \begin{matrix}
		0 & \ldots  & {{\delta }_{1}}{{x}_{0,{{\mathbb{T}}_{1}}}} & {} & 0 & \ldots  & 0  \\
		{} & {} & {} & \ddots  & {} & {} & {}  \\
		0 & \ldots  & 0 & {} & {{\delta }_{n}}{{x}_{0,{{\mathbb{T}}_{n}}}} & \ldots  & 0  \\
	\end{matrix} \right].$$
	Let
	$${{\mathbf{F}}_{\mathbb{T}}}=\left[ \begin{matrix}
		{{\delta }_{1}}{{x}_{0,{{\mathbb{T}}_{1}}}} & {} & {}  \\
		{} & \ddots  & {}  \\
		{} & {} & {{\delta }_{n}}{{x}_{0,{{\mathbb{T}}_{n}}}}  \\
	\end{matrix} \right].$$
	Then, for the whole system we have
	$$\mathbf{\tilde{L}}_n=\left[ \begin{matrix}
		{{\mathbf{G}}_{{\bar{\mathbb{T}}}}}{{\mathbf{X}}_{0,\bar{\mathbb{T}}}} & {{\mathbf{G}}_{\mathbb{T}}}{{\mathbf{X}}_{0,\mathbb{T}}}  \\
		{} & {{\mathbf{F}}_{\mathbb{T}}}  \\
	\end{matrix} \right],$$ 
	where $\mathbf{\tilde{L}}_n$ consists of $m$ rows in $\mathbf{L}_n$. And ${{\mathbf{X}}_{0,\bar{\mathbb{T}}}}$, ${{\mathbf{X}}_{0,\mathbb{T}}}$ are the cotree part and the tree part of ${{\mathbf{X}}_{0}}$, respectively. Then $rank\left( {\mathbf{\tilde{L}}_n} \right)=m$, i.e., $rank\left( \mathbf{L}_n \right)=m$.
	
	Necessity: The contrapositive of that ${{\mathbb{K}}_{D}}$ needs to cover at least one spanning tree for a complete system to detect any attack vector is, an undetectable FDI attack can always be constructed when ${{\mathbb{K}}_{D}}$ does not cover any spanning tree. 
	For a spanning tree $\mathbb{T}$, if there is a line ${{\mathbb{T}}_{l}}$ in $\mathbb{T}$ that satisfies ${{\mathbb{T}}_{l}}\notin {{\mathbb{K}}_{D}}$, then ${{\mathbb{T}}_{l}}$ and $\bar{\mathbb{T}}$ can form a cut set ${\mathbb{S}_{l}}$ of the system. When the attack set ${{\mathbb{K}}_{A}}$ is ${\mathbb{S}_{l}}$, an undetectable FDI attack can be constructed, which proves the necessity.
\end{IEEEproof} 

The deployment condition of an incomplete system can be obtained based on the results of Theorem \ref{Theorem:deployment}. Since the single-line cuts in the system must belong to a spanning tree, while the MTD on the single-line cuts is invalid, the deployment condition of the incomplete system is derived by 
\begin{equation}
{{\mathbb{K}}_{D}}={{\mathbb{T}}}\backslash {{\mathbb{K}}_{sc}}
\end{equation} 
The deployment condition of any system can be summarized as Corollary \ref{corollary:deployment}. 
\begin{corollary}\label{corollary:deployment}
	The deployment condition of any system to reach the topology-determined supremum is: ${{\mathbb{K}}_{D}}={{\mathbb{T}}}\backslash {{\mathbb{K}}_{sc}}$. Specially, ${{\mathbb{K}}_{sc}}\text{=}\varnothing $ for a complete system.
\end{corollary}

\subsection{Detection Capability's Supremum under Incomplete D-FACTS Deployment}

In this section, we will discuss the maximum detection capability  for a given D-FACTS deployment that may not satisfy Corollary \ref{corollary:deployment}.

According to Corollary \ref{corollary:deployment}, we can find that the full deployment and the deployment of ${{\mathbb{K}}_{D}}={{\mathbb{T}}}\backslash {{\mathbb{K}}_{sc}}$ have the same maximum detection capability in any system. Moreover, when the detection capability reaches this supremum, the rank of the composite matrix $ \mathbf{L} $ is \textcolor{black}{increased by} $n-{{m}_{sc}}$ compared to the initial ${{\mathbf{F}}_{0}}$. Through the proof of Theorem \ref{Theorem:deployment}, we can find that $n-{{m}_{sc}}$ is the number of elements in a maximal linearly independent group without any single-line cut for the system. This conclusion can be extended to any deployment scenario as Corollary \ref{corollary: maxline}.
\begin{corollary}\label{corollary: maxline}
	The maximum number of lines that can improve the detection capability is the largest linearly independent group excluding the single-line cuts in all deployed lines.
\end{corollary}

Let ${{{\mathbb{K}}'}_{D}}$ be a certain deployment scheme, which may contain multiple connected components. Find the maximum linearly independent group ${{\mathbb{T}}}_{D}$ in ${{{\mathbb{K}}'}_{D}}$, i.e., a forest of ${{{\mathbb{K}}'}_{D}}$ \cite{diestel2017basics}.  Then, remove the single line cuts ${{\mathbb{K}}_{sc\text{ }\!\!\_\!\!\text{ }D}}$ in ${{\mathbb{T}}}_{D}$. Assume the number of elements in ${{\mathbb{T}}}_{D}$ and ${{\mathbb{K}}_{sc\text{ }\!\!\_\!\!\text{ }D}}$ as ${{m}_{D}}$ and ${{m}_{sc\_D}}$,  respectively. Corollary \ref{key} is derived as follows.
\begin{corollary}\label{key}
	The condition to achieve the detection capability supremum under incomplete D-FACTS deployment is $ rank(\textbf{L}) = m-n+{{m}_{D}}-{{m}_{sc\_D}}. $
	
\end{corollary}

Obviously, when D-FACTS deployment satisfies Corollary \ref{corollary:deployment}, we have
\begin{equation}
    m-n+{{m}_{D}}-{{m}_{sc\_D}}=m-m_{sc}
\end{equation}  
So Corollary \ref{key} can be used to describe the supremum of detection capability under any topology and D-FACTS deployment.

\section{Implementation of MMTD}\label{sec:implementation}

\begin{figure}
 \vspace{-10pt}
	\centering
	\includegraphics[width=1\linewidth]{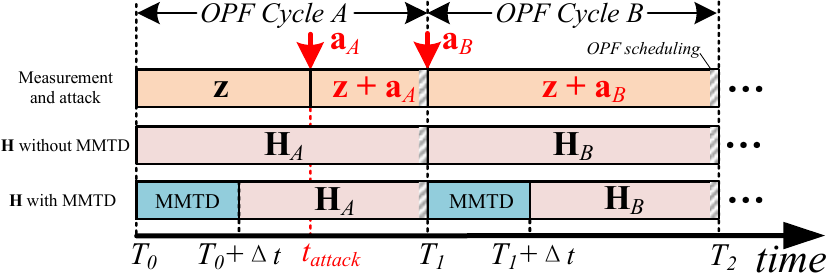}\vspace{-5pt}
	\caption{\textcolor{black}{MMTD implementation within OPF cycles.}}
	\label{fig:model}
	\vspace{-15pt}
\end{figure}

\subsection{{System} Model with MMTD}
This section proposes the system model with MMTD and OPF. {\blueEnv{}
Considering the fluctuation of system loads, the power grid performs the OPF scheduling every 5-10 minutes to work under an economic optimal condition \cite{liu2020optimal}. When D-FACTS devices are deployed in the system, they will be tuned during the OPF scheduling interval to further reduce power losses. Fig.~\ref{fig:model} shows two OPF cycles ($T_0$-$T_1$ and $T_1$-$T_2$), where the OPF scheduling occurs at the very end of each cycle.  Specifically, for traditional OPF cycles (i.e., the second line annotated as ``H without MMTD"), D-FACTS devices are tuned at $T_1$ so that the measurement matrix switches from $\mathbf{H}_A$ to $\mathbf{H}_B$.
When the MMTD strategy is applied, it will occur at the beginning of each cycle (i.e., just after the last OPF scheduling), and last for $\Delta t$ to perform a group of MTD schemes for better detection capability.

The FDI attack can occur at any time within the OPF cycles. However, frequent data modification could make the attack easier to detect. Thus, attackers must keep the attacked measurement data unchanged for a certain period. Assume that attackers know the results of OPF, a choice for better stealthiness is to persist the attacked measurement data and change them only when the measurement matrix changes. For example, as shown in Fig.~\ref{fig:model}, attackers start to attack with $\mathbf{a}_A$ at $t_{attack}$ and adjust to $\mathbf{a}_B$ at $T_1$.

When there is no MMTD in the system, as attackers know
$ \mathbf{H}_A $ and $ \mathbf{H}_B $ in advance, the FDI attack will always be undetectable. When MMTD is launched, the attack $\mathbf{a}_B$ will be detected by the MMTD in ($T_1$, $T_1 + \Delta  t$). Therefore, whenever the attack occurs, it will be detected by MMTD in the next OPF cycle.
}

\subsection{Search Algorithm for MMTD Strategy}

\begin{figure}[!t] \vspace{-18pt}
	\renewcommand{\algorithmicrequire}{\textbf{Input:}}
	\renewcommand{\algorithmicensure}{\textbf{Output:}}
	\begin{algorithm}[H]
		\caption{MMTD Search Algorithm} 
		\begin{algorithmic}[1]
			\REQUIRE  $\mathbf{x}_0$, ${{\mathbb{K}}_{sc}}$, ${{\mathbb{K}}_{D}}$, ${{\mathbf{G}}}$, $ m $, $ n $
			\ENSURE  $\mathbf{Y}$
			\STATE Initialization:    $ k=1 $;  $ \mathbf{Y}=\mathbf{x}_0 $
			\STATE $ \mathbf{X}_0=diag(x_{0,1},x_{0,2},\ldots,x_{0,m}) $; $ \mathbf{F}_0 = \mathbf{G}\mathbf{X}_0 $; $\mathbf{L} = \mathbf{F}_0$
			\STATE Calculate $ m_D $, $ m_{sc\_D} $
			\STATE ${{L}_{D}}=m-n+{{m}_{D}}-{{m}_{sc\_D}}$
			\WHILE {$ rank(\textbf{L})<L_D $}
			\STATE Generate an $ \mathbf{x}_k $ that is linearly independent of each column of \textbf{Y}
			\STATE $ \mathbf{X}_k=diag(x_{k,1},x_{k,2},\ldots,x_{k,m}) $; $ \mathbf{F}_k = \mathbf{G}\mathbf{X}_k $
			\STATE $ \mathbf{L} $ = [$ \mathbf{L} $; $ \mathbf{F}_k $]; $ \mathbf{Y} $ = [$ \mathbf{Y} $  $ \mathbf{x}_k $]; $  k = k+1 $
			\ENDWHILE
			\STATE Return $ \mathbf{Y} $.\label{algorithm:1}
		\end{algorithmic}
	\end{algorithm} \vspace{-20pt}
\end{figure}

This section discusses the search algorithm for MMTD to achieve the maximum detection capability.

There are several optimization objectives for the implementation of MMTD, such as the minimum impact on the system in the process, the fastest process, the least perturbed lines and so on. In the attack and defense scenario of MMTD, the later the system achieves the maximum detection capability, the greater the losses caused by the attack. Thus, this paper takes the least MTD stages as the optimization goal. Let $ L_D $ represent the maximum rank of $ \mathbf{L} $, i.e., $ L_D = m-n+{{m}_{D}}-{{m}_{sc\_D}}$. The optimization model is formulated as
\begin{align}\label{min_k}
	& \min k \\ 
	& s.t. ~rank({{\mathbf{L}}_{k}})={{L}_{D}}  \notag
\end{align}

(\ref{min_k}) is a rank-constrained optimization problem which has been proven to be an NP-hard problem \cite{sun2017rank}. Therefore, this paper proposes a heuristic algorithm to solve this problem. According to Lemma \ref{lemma:rank}, the effect of each MTD scheme in MMTD can be characterized by the change of $ rank(\mathbf{L}) $. For the maximum rank of $ \mathbf{L} $ is determined, an effective way to minimize $ k $ is to maximize the increase of $ rank(\mathbf{L}) $ after each MTD scheme. An approximate solution of the NP-hard problem can be got through solving
\begin{align}\label{rank+}
	&\underset{{{\mathbf{F}}_{k}}}{\mathop{\max }}\,rank\left( {{\mathbf{L}}_{k}} \right)-rank\left( {{\mathbf{L}}_{k-1}} \right)\\
	& s.t.~rank({{\mathbf{L}}_{k-1}}) < {{L}_{D}}  \notag
\end{align}

\textcolor{black}{According to Corollary \ref{corollary: maxline}, not all the deployed lines can improve the security performance. For a branch in a loop, its detection function can be replaced by other branches in this loop in the MMTD process. In other words, perturbing all deployed lines will at best produce redundant perturbations without deteriorating the final result.}

The simplest solution to reach the supremum of detection capability most quickly is to perturb all deployed lines every time. Therefore, to make the strategy more concise, we assume that all deployment lines are deemed necessary and required to be perturbed. We involve all deployment lines in the implementation scheme every time. A simple strategy can be obtained that all the deployed lines randomly generate a reactance vector $ \mathbf{x}_k $ which satisfies\begin{equation}\label{eqa.p_constrain}
	 \left( 1-\tau  \right){{x}_{0,l}}\le {{x}_{k,l}}\le \left( 1+\tau  \right){{x}_{0,l}},~l\in {{\mathbb{K}}_{D}}
\end{equation} 
where  $\tau $ is the physical limit of D-FACTS (usually $20\%$ \cite{4039419}). Specially, to ensure effectiveness, the new vector of line reactance generated after each MTD should be linearly independent of all previous reactance vectors.

To sum up, the algorithm of MTD implementation scheme is shown as Algorithm 1. The $ \mathbf{x}_k $ in step 6 must satisfy (\ref{eqa.p_constrain}). \textbf{Y} is the complete solution obtained after the entire MMTD strategy.

\begin{figure}[t] \vspace{-10pt}
	\centering
	\includegraphics[width=0.5\linewidth]{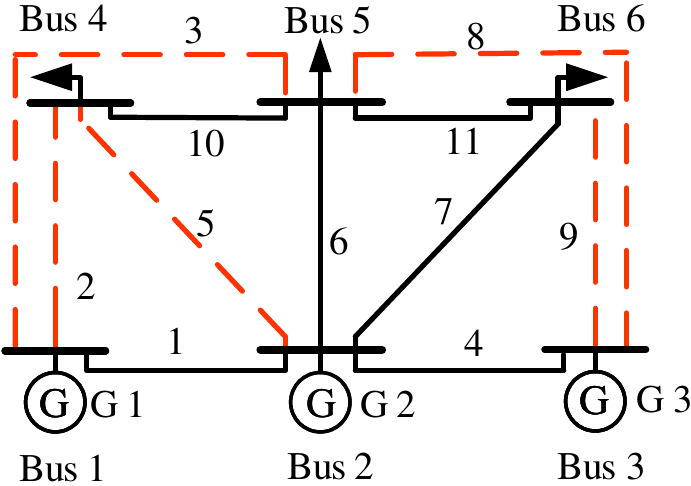} \vspace{-5pt}
	\caption{\textcolor{black}{D-FACTS deployment of the 6-bus system. The red dotted lines are the D-FACTS deployment lines.}}
	\label{fig:6busD}
\end{figure} 

\begin{figure}[t]
	\centering
	\includegraphics[width=0.85\linewidth]{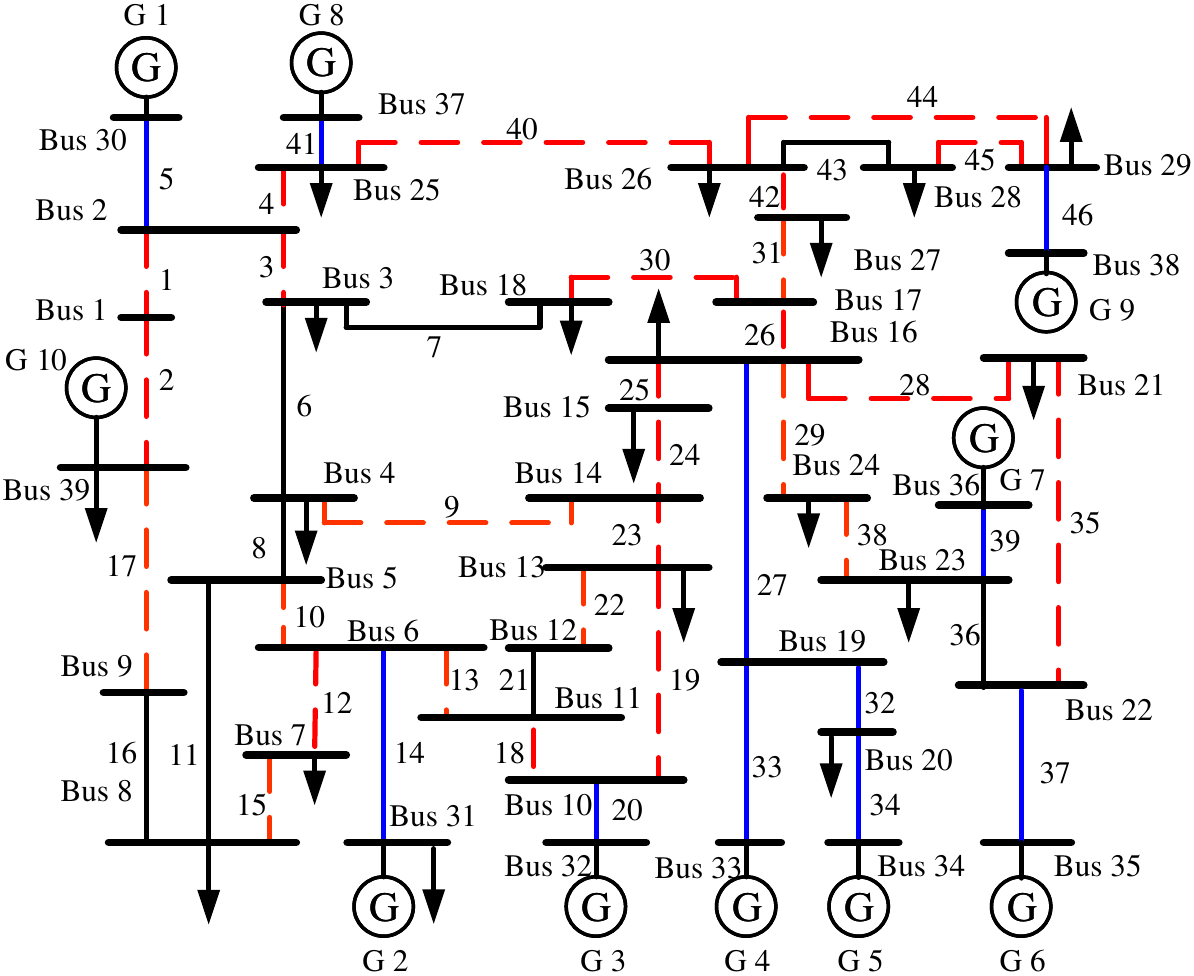}\vspace{-5pt}
	\caption{\textcolor{black}{D-FACTS deployment of the 39-bus system. The blue lines are the single-line cuts in the system.}}
	\label{fig:39busD}\vspace{-15pt}
\end{figure}

\section{Numerical Simulation and Discussion}\label{sec:simulation}

\subsection{System Setup}

To evaluate the effectiveness of the proposed MMTD strategy, we carry out experiments on IEEE 6-bus, IEEE 14-bus,
IEEE 39-bus, IEEE 57-bus, and IEEE 118-bus systems. The data of the test systems are obtained from the MATPOWER software package\cite{zimmerman2010matpower}. The SE, BDD, FDI attacks and MMTD strategy are written in MATLAB and executed on a desktop computer with Intel Core-i7 CPU 2.70 GHz and 16GB RAM. The experimental control groups of this paper are set as follows:

\textbf{ECASE}: The economic-oriented perturbation scheme. The ECASE strategy consistent with \cite{rogers2008some} can be derived by
\begin{align}
	& \min {{P}_{loss}} \\ 
	& s.t.~\left( 1-\tau  \right){{x}_{0,l}}\le {{x}_{l}}\le \left( 1+\tau  \right){{x}_{0.l}},~l\in {{\mathbb{K}}_{D}} \notag 
\end{align}
where ${{P}_{loss}}$ is the real power losses of the system. $\tau $ is set to 20\% \cite{4039419}.

\textbf{RMTD}: Random MTD strategy. The earliest MTD which is to carry out random perturbation on random lines in the system. The RMTD has randomness in both deployment and implementation \cite{rahman2014moving}.

\textbf{CCPA}: Based on the rank-maximizing strategy, $ m-n $ D-FACTS are deployed in the system, and $rank(\left[ \begin{matrix}
	\mathbf{H}_0 & {\mathbf{{H}_1}}  \\
\end{matrix} \right])  $  after MTD is used to measure MTD's detection capability. The aim of this strategy is to achieve the maximum rank with the least D-FACTS devices \cite{lakshminarayana2019moving}.

\textbf{PFDD}: This study proposes a spanning tree deployment strategy in the system \cite{li2019feasibility}.

\textbf{SRPA}: This study determines the direction for all lines to reduce the power losses to solve the joint optimization problem of economic and security objectives of MTD \cite{liu2018reactance}.
\subsection{D-FACTS Deployment Solutions}

\begin{table}[t] \vspace{-10pt}
	\renewcommand{\arraystretch}{1.3}
	\centering
	\caption{D-FACTS Deployment Number of Different MTD Strategies in Five  Test Systems}
	\label{tab:Dnumber}
	\begin{tabular}{c c c c c c}
		\toprule
		 Test system & ECASE& CCPA & SRPA & PFDD & MMTD \\ \midrule
		6-bus system   & 6    & 6    & 5    & 5    & 5    \\ \hline
		14-bus system   & 7   & 7    & 7    & 13   & 12   \\ \hline
		39-bus system   & 24   & 8    & 8    & 38   & 27   \\ \hline
		57-bus system    & 14  & 24   & 24   & 56   & 55   \\ \hline
		118-bus system   & 70  & 69   & 69   & 117  & 108  \\ \bottomrule 
	\end{tabular}\vspace{-5pt}
\end{table} 

\begin{figure}[t]
	\centering
	\includegraphics[width=0.7\linewidth]{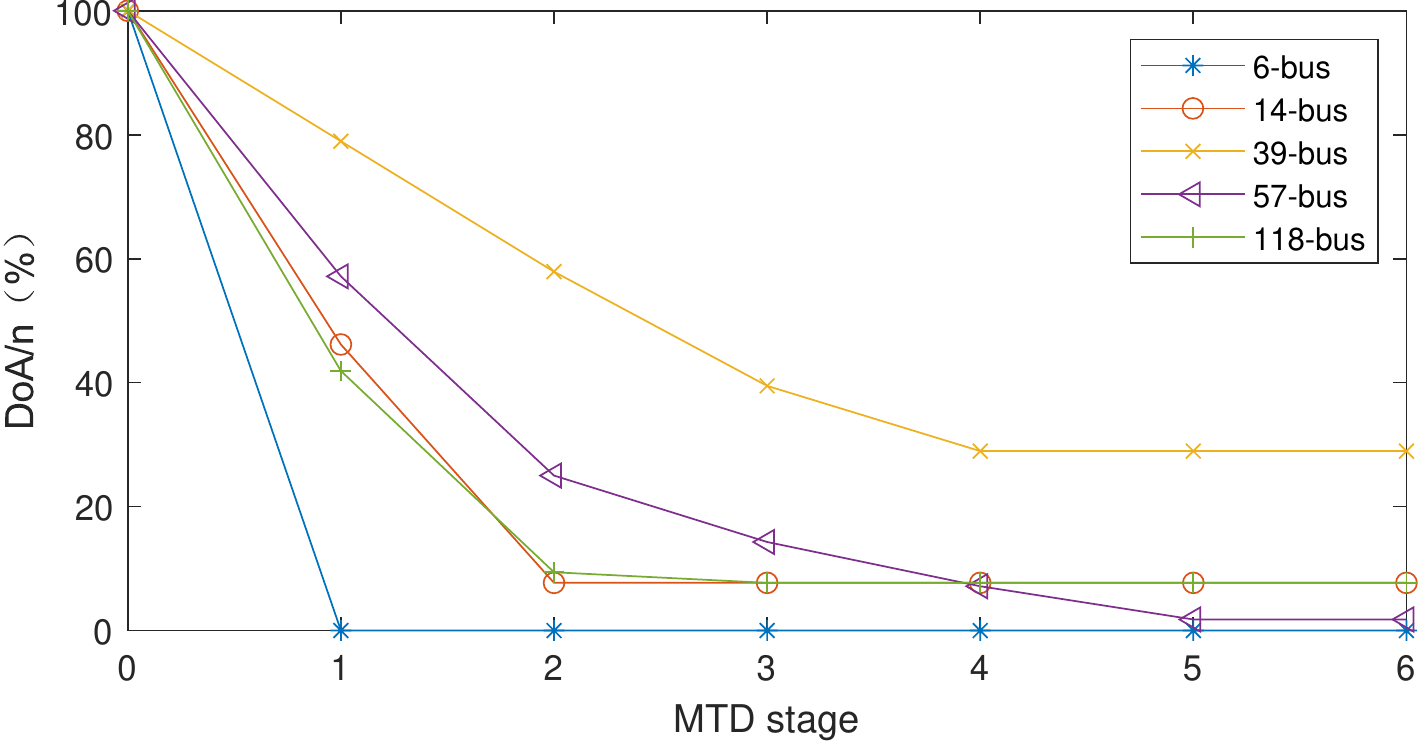}\vspace{-5pt}
	\caption{\textit{DoA}/$ n $ versus MTD stages in five test systems.}
	\label{fig:doa}\vspace{-15pt}
\end{figure}
This section compares the required number of D-FACTS devices in different MTD \blueText{strategies} and gives deployment examples of MMTD in several IEEE test systems.

In terms of the number of deployments, RMTD is not included in this comparison because of the randomness of its deployment and implementation scheme. For the deployment solution of ECASE, we deploy D-FACTS on the lines that reduce the power losses exceeding 0.1 MW. The deployment numbers of CCPA \cite{lakshminarayana2019moving}, SRPA \cite{liu2018reactance}, and  PFDD  \cite{li2019feasibility} are $ m-n $, $\min \{ m-n,n \}$, and $ n $, respectively. The  deployment number of MMTD strategy is $n-{{m}_{sc}}$. As shown in Table \ref{tab:Dnumber}, the deployment number of MMTD is moderate among all the 
strategies. Next, we perform the deployment schemes in the 6-bus and 39-bus systems, respectively. \textcolor{black}{Based on Corollary \ref{corollary:deployment}, we can derive a deployment solution if we know the line weight of the system \cite{liu2020optimal}. The weights are determined by different issues, such as loss minimization and voltage control \cite{rogers2008some}. This paper considers the weight under the target of minimizing real power losses.
As shown in Fig. \ref{fig:6busD} and Fig. \ref{fig:39busD}, the red dotted lines are the D-FACTS deployment lines, and the blue lines are the single-line cuts in these systems. From Fig. \ref{fig:6busD}, complete detection can be achieved on the 6-bus system because there is no single-line cut in this system. The deployment scheme is a spanning tree of the system, which can be \{2, 3, 5, 8, 9\}. From Fig. \ref{fig:39busD}, line \{5, 14, 20, 27, 32, 33, 34, 37, 39, 41, 46\} are single-line cuts, and the MTD on these lines are invalid. Therefore, the deployment solution of the 39-bus system is \{1, 2, 3, 4, 9, 10, 12, 13, 15, 17, 18, 19, 22, 23, 24, 25, 26, 28, 29, 30, 31, 35, 38, 40, 42, 44, 45\} without the single-line cuts in the system.}

\subsection{Comparison of DoA/n of Different MTD \blueText{Strategies}}

This section compares the attack space after each MTD \blueText{stage} to illustrate the change of detection capabilities.

As mentioned in Section \ref{sec:MMTD}, \textit{DoA}/$ n $ can be used to evaluate the detection capability of MTD. The smaller the \textit{DoA}/$ n $, the higher the detection capability of MTD. All control groups in this paper are one-stage MTDs. Fig. \ref{fig:doa} shows the change curve of \textit{DoA}/$ n $ with the number of MTD stages in 5 test systems. The MTD scheme for each stage of MMTD is derived from Algorithm 1. As shown in Fig. \ref{fig:doa}, \textit{DoA}/$ n $ of 6-bus system is zero after one MTD stage, for the 6-bus system is a complete system which has no single-line cut and satisfies   $m\ge 2n$. The minimum \textit{DoA}/$ n $ of \blueText{the} 14-bus system can be obtained after two MTD stages. \textcolor{black}{The minimum \textit{DoA}/$ n $ of the  39-bus system is the largest among five test systems, which has the highest proportion of single-line cuts. The \textit{DoA}/$ n $ of the 57-bus system reaches the lower bound after five MTD stages, while the 118-bus system needs three stages to reach the lower bound. In general, the final results of MMTD in all incomplete systems are much smaller than that of all one-stage MTDs.}

\subsection{Attack Detection Probabilities of Different MTD Strategies}
\begin{figure}[t]\vspace{-10pt}
	\centering
	\includegraphics[width=0.7\linewidth]{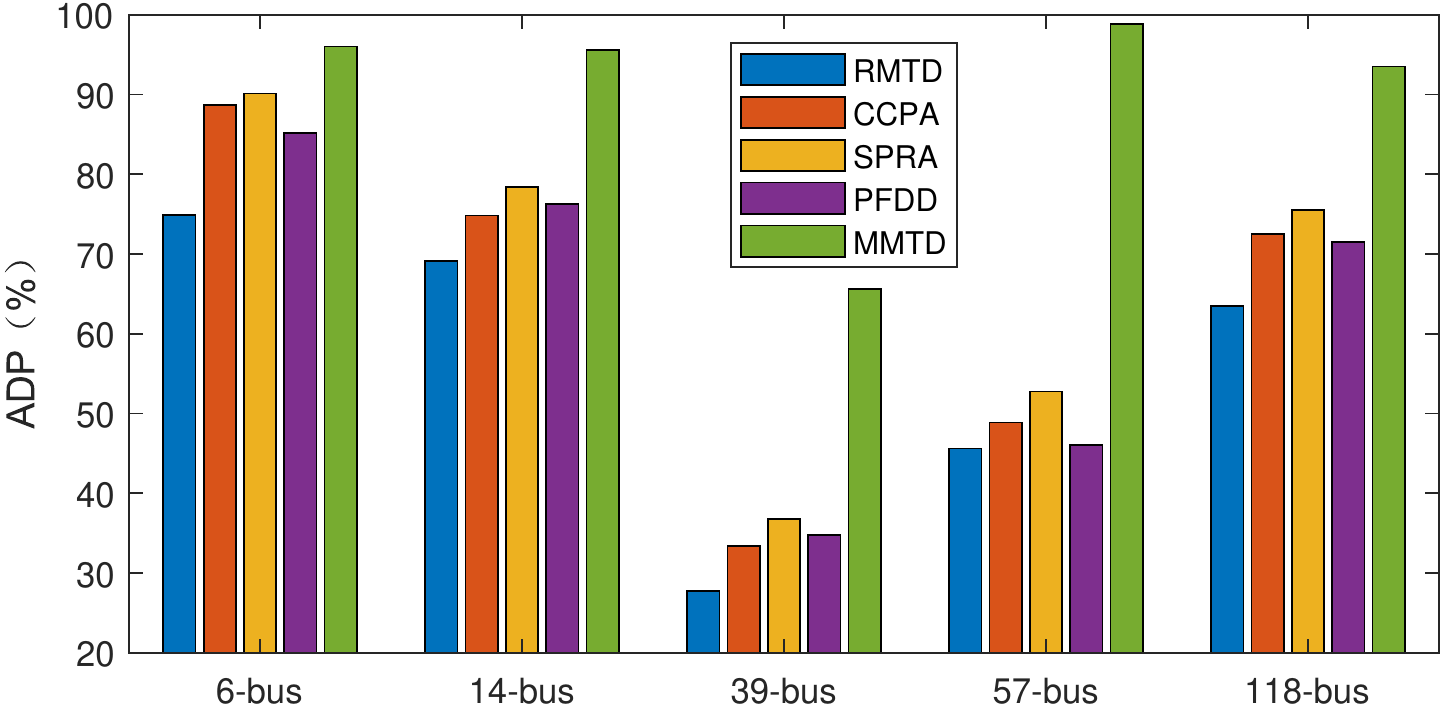}\vspace{-5pt}
	\caption{\textcolor{black}{ADP comparison with measurement noises in DC-SE.}}
	\label{fig:dc}\vspace{-10pt}
\end{figure}

\begin{figure}[t]
	\centering
	\includegraphics[width=0.7\linewidth]{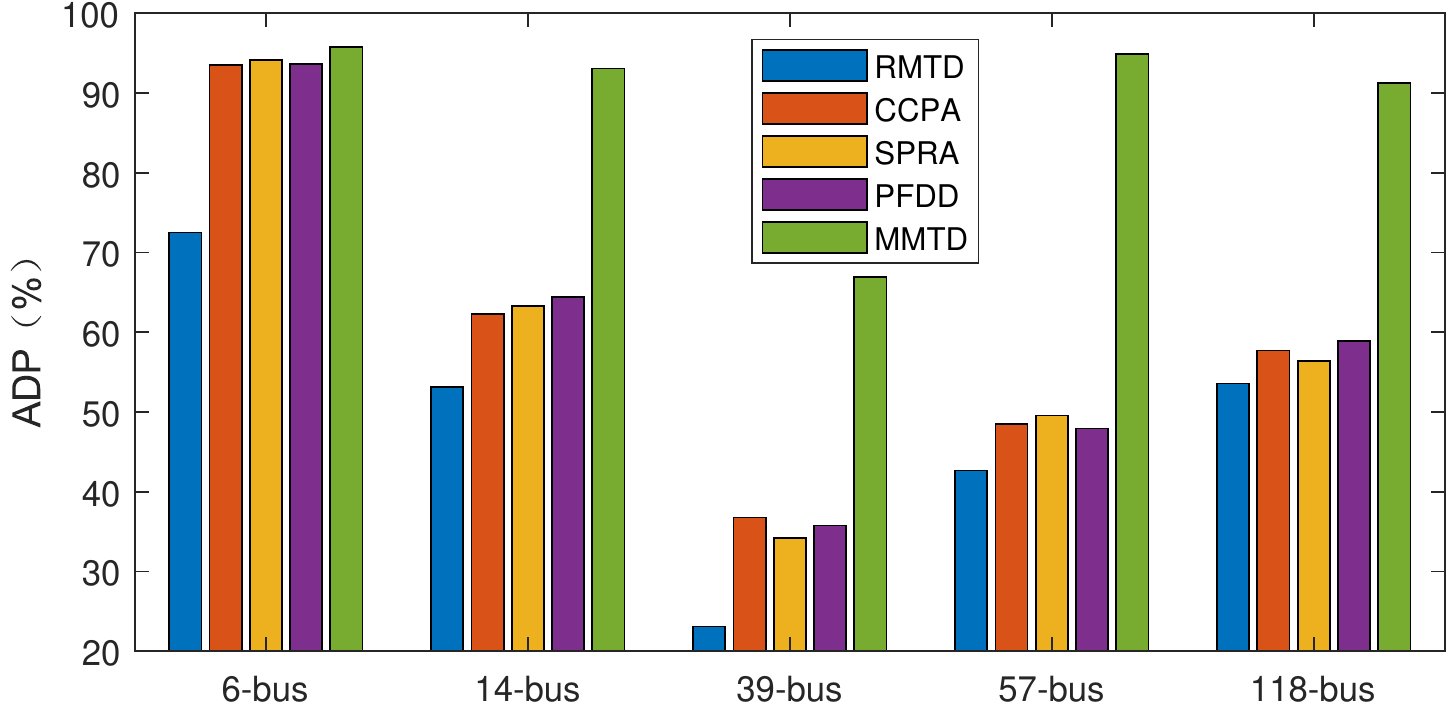}\vspace{-5pt}
	\caption{\textcolor{black}{ADP comparison with measurement noises in AC-SE.}}
	\label{fig:ac}\vspace{-15pt}
\end{figure}

\begin{table}[t]\vspace{-10pt}
	\renewcommand{\arraystretch}{1.1}
	\centering
	\caption{\textcolor{black}{Detection Rate of Each Stage within an MMTD Scheme in IEEE 57-bus System}}\vspace{-5pt}
	\begin{tabular}{c c c c c c c}
		\toprule
		Stage & 1   & 2  & 3   & 4  & 5  & Total \\ \midrule
		\makecell{Detection\\ rate (\%)} & 50.19 & 51.6 & 46.85 & 48.11 & 47.67 & 98.65\\ 
		 \bottomrule
	\end{tabular}
	\label{tab:everystage}\vspace{-15pt}
\end{table}

In this section, we conduct comparative experiments of attack detection probability (ADP) of different MTD strategies. \textcolor{black}{The experiments are conducted in the noisy environment in both DC-SE and AC-SE. In each set of experiments, we carry out 10000 random attack experiments. In the DC model, all the attack vectors satisfy $\mathbf{a}={{\mathbf{H}}_{0}}\mathbf{c}$, which are randomly generated in the column space of the initial measurement matrix $\mathbf{H}_0$. In particular, all attacks are carried out on only one basic cut of the system to make the attack harder to detect. The measurement noise is assumed to be Gaussian distributed with zero mean and the standard deviation is set as 1\% of the actual measurement \cite{liu2018reactance}. BDD is used to detect the FDI attacks, and its threshold is set at a false positive rate of $\alpha \text{=0}\text{.05}$. To ensure the detection efficiency, the perturbation of line reactance in the noisy environment should be greater than 5\% \cite{liu2018reactance}.}

\textcolor{black}{As shown in Fig. \ref{fig:dc}, the ADP of MMTD is significantly ahead of other MTD strategies in the five test systems. Among the five test systems, only the 6-bus system is a complete system, in which the minimum \textit{DoA}/$ n $ is zero. All MTD strategies perform well in the 6-bus system. The ADP of all MTD strategies in the 39-bus system is lower than other systems. This is because 11 of the 46 lines in the 39-bus system are single-line cuts. In all test systems, the relationship between the ADP and the proportion of single-line cuts is consistent, where a higher proportion of single-line cuts generally corresponds to a lower detection probability.}

\textcolor{black}{Further, we test the effectiveness of MMTD in the AC model. Different from the DC model, the attacker needs to know system states, such as voltage magnitude and angle, to construct AC attacks. The AC-FDI attack can be constructed as $\mathbf{a}=\mathbf{h}(\hat{\boldsymbol{\thetaup}}+\Delta\boldsymbol{\thetaup})-\mathbf{h}(\hat{\boldsymbol{\thetaup}})$, where $\Delta\boldsymbol{\thetaup}$ is the state variable determined by the attacker, and $\hat{\boldsymbol{\thetaup}}$ is the system state estimated by the attacker. The MMTD strategy in the AC model is the same as that in the DC model. The final ADP results are shown in Fig. \ref{fig:ac}. From Fig. \ref{fig:dc} and Fig. \ref{fig:ac}, the detection effectiveness of MMTD in the AC model is consistent with that in the DC model, where the MMTD strategy has a leading detection probability in each system.}

\textcolor{black}{In addition, we take the 57-bus system with five stages as an example to illustrate the effect of MMTD on improving the detection capability. The stage 1-stage 5 in Table \ref{tab:everystage} are five sub-schemes of an MMTD scheme. All these five sub-schemes have the maximum detection capability that a one-stage MTD can have. However, the detection rates of these sub-schemes are all less than 60\% due to topology constraints. The last column shows that the detection rate of the overall MMTD scheme is 98.65\%. This is because subsequent sub-schemes can detect attacks that can bypass previous sub-schemes.}

\subsection{Economic Performance Analysis}

\begin{figure}[t]
	\centering
	\includegraphics[width=0.65\linewidth]{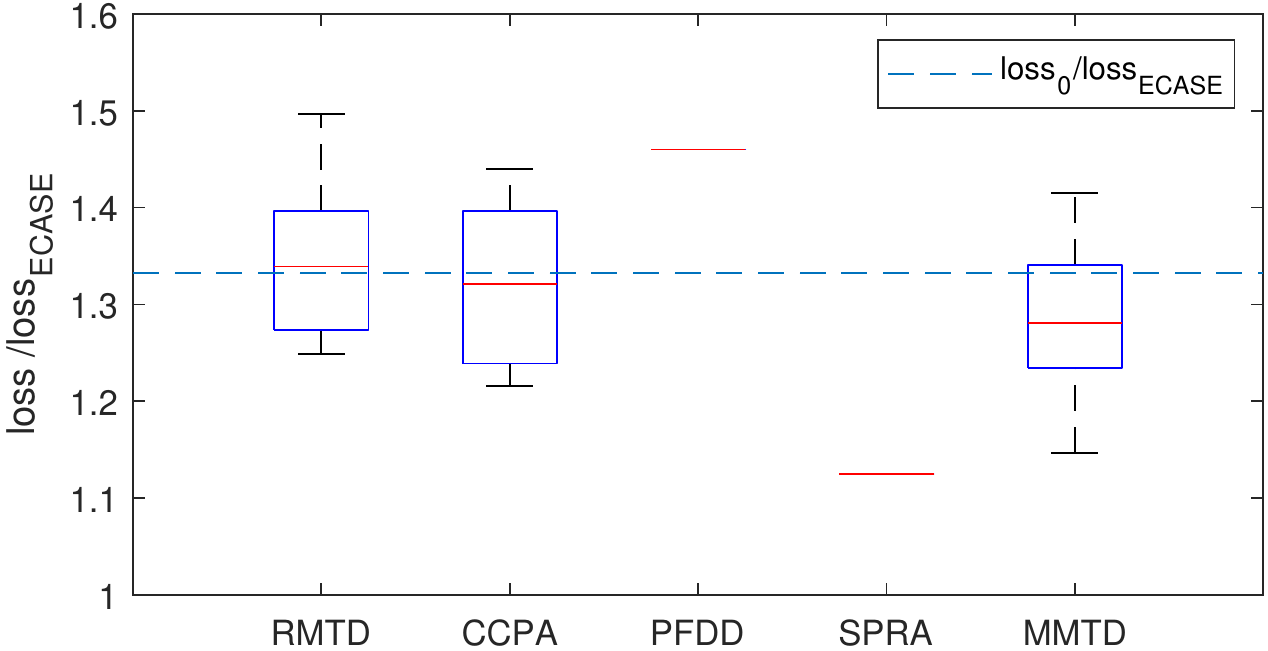}\vspace{-5pt}
	\caption{Economic performance comparison in the 57-bus system.}
	\label{fig:erate}
 	\vspace{-5pt}
\end{figure}

This section compares the economic performance of different MTD strategies. Fig. \ref{fig:erate} shows the power losses comparison of each MTD strategy in the 57-bus system, where the y-axis is the ratio of the power losses of different MTD strategies to ECASE. The blue dotted line is the ratio of the power losses before MTD to ECASE, which is 1.3327. The power losses ratios of each MTD strategy after ten experiments are shown in a boxplot, where each result of MMTD is the average of the five stages of that time. All ten results of PFDD and SPRA are the same, for these two strategies do not contain randomness.

\textcolor{black}{According to the results in Fig. \ref{fig:erate}, we then discuss the economic performance of a whole OPF cycle of power system, i.e., the $ T_0 - T_1 $ stage in Fig. \ref{fig:model}. We ignore the OPF scheduling in the cycle because its time is very short, and divide a cycle into the perturbation stage ($T_0$, $T_0 + \Delta  t$) and the steady stage ($T_0 + \Delta  t$, $T_1$). Suppose the whole OPF cycle in Section \ref{sec:implementation} is five minutes, and the perturbation stage with five stages is 25 seconds. The average power losses of the whole OPF cycle can be obtained by weighting the average power losses of each stage.} All the one-stage MTD strategies will replace the original economic-oriented scheme of the power system, While the MMTD strategy will return to the economic-oriented scheme after the perturbation stage.  According to the results of each MTD strategy in the whole OPF cycle in Table \ref{tab:erate}, the MMTD strategy proposed in this paper is smaller than the other several MTD strategies.

\begin{table}[t]
	\renewcommand{\arraystretch}{1.3}
	\centering
	\caption{Average Power Losses Ratio of Different MTD Strategies at Different Stages}\vspace{-5pt}
	\begin{tabular}{c c c c c c}
		\toprule
		System stage& RMTD   & CCPA   & PFDD   & SPRA  & MMTD   \\ \midrule
		Perturbation stage & 1.3513 & 1.3232 & 1.4599 & 1.125 & 1.2866 \\ \hline
		\specialrule{0em}{0pt}{0.5pt} Steady stage     & 1.3513 & 1.3232 & 1.4599 & 1.125 & 1      \\ \hline
		 Whole OPF cycle & 1.3513 & 1.3232 & 1.4599 & 1.125 & 1.0478 \\ \bottomrule
	\end{tabular}
	\label{tab:erate}\vspace{-8pt}
\end{table}
\section{Conclusion and Future  Work}\label{sec:conclusion}
In this paper, we propose an MMTD strategy to solve the conflict between MTD's security goals and D-FACTS' economic goals, and maximize the system's detection capability. We propose a metric for MTD's detection capability, find its supremum for any topology and D-FACTS deployment, and design an algorithm to reach this supremum. Finally, we verify the superiority of the MMTD strategy in detecting FDI attacks through extensive experiments. Simulation results indicate that the MMTD strategy could detect FDI attacks with admirable accuracy while causing rather low power losses during the OPF cycle.

 \textcolor{black}{
 In the system model proposed in Section~\ref{sec:implementation}, there is a delay in detection because the economic-oriented scheme has no detection capability. Moreover, the attacker may detect the power flow change generated by existing MMTD schemes. To address these issues, we will work on reducing the detection delay and study the hiddenness of MMTD in future work.}

\ifCLASSOPTIONcaptionsoff
  \newpage
\fi



%
\bibliographystyle{./IEEEtran}
\bibliography{mybib.bib}
%
%

%

%
%
%




\end{document}